\documentclass{ws-ijmpa}
\catcode`@=11 %This allows us to modify plain macros
\def\as{\alpha_s}
\def\lsim{\mathrel{\mathpalette\@versim<}}
\def\gsim{\mathrel{\mathpalette\@versim>}}
 \def\@versim#1#2{\lower0.2ex\vbox{\baselineskip\z@skip\lineskip\z@skip
       \lineskiplimit\z@\ialign{$\m@th#1\hfil##$\crcr#2\crcr\sim\crcr}}}
\catcode`@=12 %at signs are no longer letters
\def\tozero#1{\mathrel{\mathop{\sim}\limits_{\scriptscriptstyle {#1\rightarrow0 }}}}

\begin{document}
%%%%%%%%%%%%%%%%%%%%%%%%%%%%%%%%%%%%%%%%%%%%%%%%%%%%%%%%%%%%%%%%%%%%%%%%%%%%%%%%%%%
%%%%%%%%%%%%%%   CUT HERE %%%%%%%%%%%%%%%%%%%%%%%%%%%%%%%%%%%%%%%%%%%%%%%%%%%%%%%%%%%%
%%%%%%%%%%%%%%%%%%%%%%%%%%%%%%%%%%%%%%%%%%%%%%%%%%%%%%%%%%%%%%%%%%%%%%%%%%%%%%%%%%%
\pagestyle{empty}
{\large
\begin{center}

{\Large\bf Progress in Small $x$ Resummation}\\
\bigskip
\bigskip
{\large\bf  Stefano Forte}\\

\bigskip
{ \it Dipartimento di Fisica, Universit\`a di Milano and\\ 
INFN, Sezione di Milano, Via Celoria 16, I-20133 Milano, Italy}
%\maketitle
\bigskip
\bigskip
\bigskip
\vskip2cm

{\bf Abstract}\\
\end{center}
\bigskip
\noindent
I review recent theoretical progress in the resummation of small $x$
contributions to the evolution  of parton distributions, in view of
its potential significance for accurate phenomenology at future
colliders. I show that a consistent perturbative 
resummation of collinear and
energy logs is now possible, and   necessary if one wishes to use
recent NNLO results in the HERA kinematic region.\hfill\\
\vskip1.6cm
\begin{center}
 Invited plenary talk at the\\
{\bf International Conference on QCD and
 Hadronic Physics}\\
Beijing, June 2005\\
{\it to be published in the proceedings}
\end{center}
\vfill
September 2005\hfill IFUM-847/FT}
%\end{flushright}
\eject
\setcounter{page}{1} \pagestyle{plain}
%%%%%%%%%%%%%%%%%%%%%%%%%%%%%%%%%%%%%%%%%%%%%%%%%%%%%%%%%%%%%%%%%%%%%%%%%%%%%%%%%%%
%%%%%%%%%%%%%%   CUT HERE %%%%%%%%%%%%%%%%%%%%%%%%%%%%%%%%%%%%%%%%%%%%%%%%%%%%%%%%%%%%
%%%%%%%%%%%%%%%%%%%%%%%%%%%%%%%%%%%%%%%%%%%%%%%%%%%%%%%%%%%%%%%%%%%%%%%%%%%%%%%%%%%

\markboth{Stefano Forte}
{Progress in small $x$ resummation}

\catchline{}{}{}

\title{PROGRESS IN SMALL $x$ RESUMMATION}

\author{\footnotesize STEFANO FORTE}

\address{
Dipartimento di  Fisica, Universit\`a di
Milano and\\INFN, Sezione di Milano, Via Celoria 16, I-20133 Milan, Italy}

\maketitle

%\pub{Received (Day Month Year)}{Revised (Day Month Year)}

\begin{abstract}
I review recent theoretical progress in the resummation of small $x$
contributions to the evolution  of parton distributions, in view of
its potential significance for accurate phenomenology at future
colliders. I show that a consistent perturbative 
resummation of collinear and
energy logs is now possible, and   necessary if one wishes to use
recent NNLO results in the HERA kinematic region.
%\keywords{Keyword1; keyword2; keyword3.}
\end{abstract}

\section{The dangerous success of NLO calculations}
There is a general attitude in the collider
physics community 
that  small $x$ resummation is impossible to understand, but there is
no need to worry since the data can be described very well without it.
It is certainly true that (for instance) standard global
parton fits~\cite{Lai:1999wy,Martin:2002aw} are based
on fixed next-to-leading (NLO)  
order calculations,
and that they manage to describe the data very accurately, in
particular throughout the HERA kinematic region.
The lack of experimental evidence for higher-order perturbative
corrections at small $x$, despite theoretical arguments which suggest
that they should be large, has been the motivation for a substantial
amount of theoretical effort over the last
decade.~\cite{Ball:1995vc}$^{-}$\cite{Thorne:2001nr}
This activity has become  less of an academic exercise since
the recent determination of the full set of NNLO splitting
functions~\cite{Moch:2004pa,Vogt:2004mw} and of the full set of
$O(\as)$ perturbative corrections to deep-inelastic
scattering.~\cite{Moch:2005xu,Vermaseren:2005qc} Indeed, parton fits
that include these NNLO
terms~\cite{Martin:2004ir,Thorne:2005da} appear 
to be unstable 
  at small $x$: when
$x \lsim 10^{-3}$ the difference in results e.g. for
  the $P_{gg}$ splitting function or for the gluon-dominated structure
  function $F_L$  when going from
NNLO and NLO starts becoming
  as large or larger than the difference between NLO and LO.
Interestingly, it also turns out that at small $x$ there is a
significant difference between the full NNLO result and the nominally
leading small $x$ contributions: subleading terms are crucial.
This means that, whatever the reason for the success of NLO fits, 
small $x$ resummation is mandatory beyond NLLO, at least for
$x<10^{-3}$.

Thanks to recent theoretical progress, there is now  a consistent
theory which enables small $x$ resummation, and which is close to
being amenable to realistic phenomenology. This theory
requires several
ingredients, which were developed by various people over the last
decade, and have recently
led to two approaches 
(ABF~\cite{Ball:1995vc}$^{-}$\cite{Altarelli:2005xx}  
and
CCSS~\cite{Salam:1998tj}$^{-}$\cite{Ciafaloni:2003rd})
which incorporate
similar basic principles, and which  arrive at stable and
consistent resummed results. Here we will mainly review  the ABF
approach, while also comparing with the results of the CCSS
approach and briefly discussing differences between the two approaches.

The main problems in small $x$ resummation
of the standard GLAP evolution equations is first, that leading
logarithmic resummation corrections are to be too large --- they would lead
to a powerlike small $x$ rise of splitting functions and thus of
structure functions which is incompatible with the data,~\cite{Ball:1995vc,Blumlein:1997em} and second,
that they are perturbatively unstable --- the next-to-leading
logarithmic  resummation corrections are larger than the leading ones.

These problems are solved thanks to three main ingredients:
duality~\cite{Ball:1997vf,Ball:1999sh,Altarelli:1999vw}, which makes
the joint resummation of collinear and small $x$ logs
possible, running coupling small $x$ resummation~\cite{Ciafaloni:1999yw,Altarelli:2001ji} and
factorization~\cite{Altarelli:2001ji}, which softens the resummed  small $x$
behaviour, and gluon exchange
symmetry~\cite{Salam:1998tj,Ciafaloni:2003rd,Altarelli:2005xx}  which stabilizes the
resummed perturbative expansion.

\section{Duality}
\begin{figure}
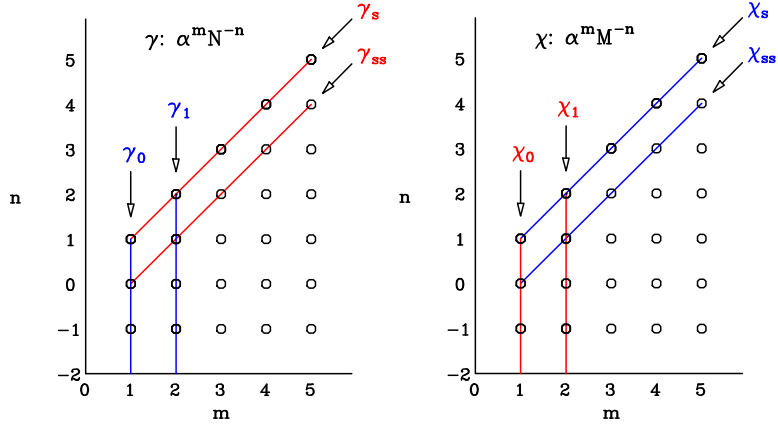

\centerline{\psfig{file=crosgam.ps,width=0.4\linewidth}
\psfig{file=croschi.ps,width=0.4\linewidth}}
\vspace*{2pt}
\caption{Dual expansions of the BFKL kernel $\chi$ and the GLAP
  anomalous dimension $\gamma$.}
\end{figure}
As is well known, singlet splitting functions at small $x$  receive
contributions of the form $\as\frac{1}{x}(\as\ln\frac{1}{x})^n( c^0_n+ \as
c^1_N+\dots)$ 
to all perturbative orders. These contributions can be extracted from
the BFKL equation, which for a  parton distribution $G(x,Q^2)$
(in fact, an eigenvector of the singlet evolution matrix), and its
Mellin transforms
\begin{eqnarray}
G(N,t)&=&\int^{\infty}_{0}\! d\xi\, e^{-N\xi}~G(\xi,t)\label{nmellin}
\\
G(\xi,M)&=&\int^{\infty}_{-\infty}\! dt\, e^{-Mt}~G(\xi,t) \label{mmellin}
\end{eqnarray}
takes the form
\begin{equation}
\frac {d}{d\xi}G(\xi,M)=\chi(M,\as)~G(\xi,M);\quad\xi\equiv \frac{1}{x},\label{bfkl}
\end{equation}
of a standard renormalization-group (or GLAP) equation,
\begin{equation}
\frac {d}{dt}G(N,t)=\gamma(N,\as)~G(N,t);\quad t\equiv \frac{Q^2}{\mu^2}
\label{rge}
\end{equation}
 but with the roles of
the variables $x$ and $Q^2$ interchanged. Upon Mellin transformation,
$\frac{1}{x}\xi^{k-1}\leftrightarrow \left(\ln
\frac{1}{N}\right)^k$ and 
$\frac{Q^2}{\mu^2}t^{k-1}\leftrightarrow \left(\ln
\frac{1}{M}\right)^k$.

\begin{figure}
\centerline{\psfig{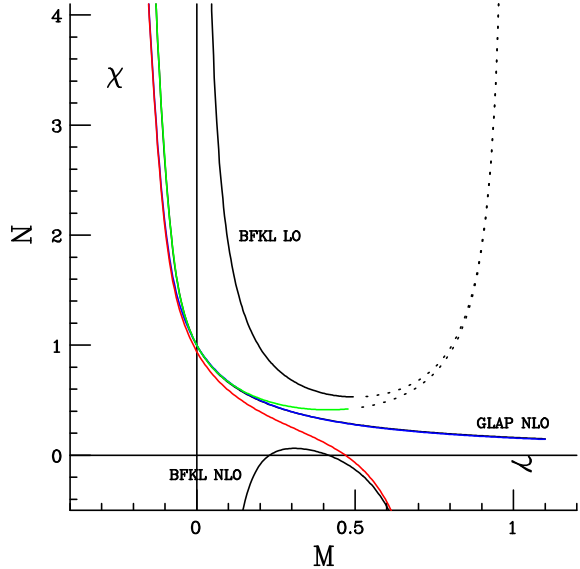}}
\vspace*{2pt}
\caption{Various expansions of the dual $\chi$ kernel, determined with
  $n_f=4$, $\as=0.2$. Note that the
  associate dual Eq.~(\ref{dual}) $\gamma(N)$
  anomalous dimension is simply the inverse function of $\chi(M)$, as per the axis
  labelling.}
\end{figure}
Whereas the way to extract the coefficients of the leading singular
$\left(\frac{\as}{N}\right)^n$ 
 contributions to the anomalous dimension
$\gamma(N)$ Eq.~(\ref{rge}) from the (BFKL) kernel $\chi(M)$ of Eq.~(\ref{bfkl})
has been known since a long time,\cite{Jaroszewicz:1982gr}
 it has been realized only more recently~\cite{Ball:1997vf,Ball:1999sh,Altarelli:1999vw} that in
fact, up to higher twist corrections, the BFKL and GLAP equations
admit the same solution, provided the boundary conditions are suitably
matched, and the corresponding kernels satisfy the duality relations
\begin{eqnarray}\chi(\gamma(N,\as),\as) &=&   N\label{dual}\\
\gamma(\chi(M,\as),\as) &=&   M. \label{revdual}
\end{eqnarray}
These relations hold at fixed coupling, whereas in the running case
they are corrected by terms which may be computed order by order in
perturbation theory. 

Duality maps the expansion of $\gamma$ in powers
of $\as$ at fixed $N$ (in Fig.~1 leading $\gamma_0$, next-to-leading
$\gamma_1$ etc.) into the expansion of $\chi$ in powers of of $\as$ at fixed $\as/M$
(in Fig.~1 leading $\chi_s$, next-to-leading
$\chi_{ss}$ etc.): hence $\gamma_0$ or $\chi_s$ sum leading logs of
$Q^2$ (collinear logs) while $\gamma_s$ or $\chi_0$ sum leading logs of
$\frac{1}{x}$ (energy logs) and so on. A joint (double-leading) resummation can be
constructed by simply combining the two expansions and subtracting the
terms which are in common: so the LO DL anomalous dimension is
$\gamma_{\rm
  DL,\,LO}(N)=\alpha_s\gamma_0(N)+\gamma_s\left(\as/N\right)-
\frac{3\alpha_s}{\pi N}$ and so on. A nontrivial
property of the DL expansion is that the dual of LO DL $\chi$ is LO DL
$\gamma$ (up to subleading terms), and similarly at next-to-leading
order and so on.

The DL expansion allows one to understand and cure the notorious
problem of the large size of subleading $\chi_1$ in comparison to
leading $\chi_0$, seen in Fig.2, where BFKL LO denotes $\as\chi_0$
and BFKL NLO denotes $\as\chi_0+\as^2\chi$. Indeed, momentum
conservation implies that to all perturbative orders $\gamma(1)=0$. By
duality this also implies $\chi(0)=1$. But $\chi$ is a polynomial in
$\as$, so $\chi(0)=1$ means that as $M\to0$ $\chi$ must behave as
\begin{equation}
\chi_s(M)\tozero M \frac{\alpha}{\alpha+M}=
\frac{\alpha}{M}-\frac{\alpha^2}{ M^2}+\frac{\alpha^3}{M^3}+\dots
\label{mom}
\end{equation}
up to subleading corrections. Hence the expansion of $\chi$ in powers of $\as$
has alternating-sign poles at $M=0$, which are resummed into $\chi_s$
and thus absent in double-leading $\chi$.
Indeed, Fig.~2 also displays the DL LO and NLO curves, which agree
respectively with LO and NLO GLAP when $M\lsim0$, and with LO and NLO
BFKL when $M\sim\frac{1}{2}$. Hence, the DL expansion is stable for
all $M\lsim\frac{1}{2}$.

\section{Running coupling}

\begin{figure}
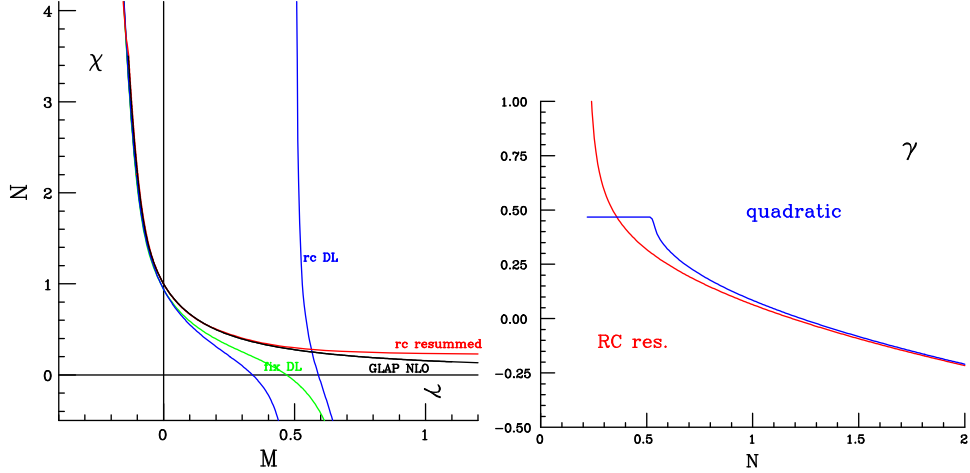

\centerline{\psfig{file=dualrcres.ps,width=0.5\linewidth}
\psfig{file=aiad.ps,width=0.5\linewidth}}
\vspace*{2pt}
\caption{Right: the singular running coupling correction to the NLO DL $\chi$
and it resummation. Left: running coupling resummation of the
anomalous dimension $\gamma$ associated to a quadratic $\chi$ kernel.}
\end{figure} The running of the coupling
$\alpha(t)=\alpha_{\mu}[1-\beta_0\alpha_{\mu} t+\dots]$
 is a  leading log $Q^2$, but next--to--leading log
$\frac{1}{x}$ effect. As a consequence, beyond LL$x$ 
the fixed-coupling duality relations~(\ref{dual}) get
 corrected: for instance $\gamma_{ss}$ 
 determined from $\chi_0$ and $\chi_1$ according to Eq.~(\ref{dual})
 must be supplemented by a running-coupling correction 
$\Delta\gamma_{ss}=-\beta_0\frac{\chi_0^{\prime\prime}\chi_0}{2(\chi_0^\prime)^2}$.
One can view these running coupling corrections as a contribution to
 an ``effective'' $\chi$ which then respects the duality
 relation~(\ref{dual}) even at the running coupling level.

The good thing about these running coupling correction is that one can
show that through their inclusion duality between the BFKL and
GLAP equations holds to all perturbative orders. This means that the
running-coupling BFKL equation, just like the GLAP equation, 
admits a factorized solution, whose
scale dependence is determined by the kernel independent of the boundary
condition.~\cite{Altarelli:2001ji}
The bad thing is that the associate effective $\chi$ is singular at
the minimum of the fixed-coupling LO $\chi$, as shown in Fig.~3.
This singularity implies that the associate
splitting function grows as a power of $\xi$ in comparison to the
leading-order one as $x\to0$: at $N^{k}LO$, $\Delta
P_{s^k}(\as,x)/P(\as,x)\tozero x (\beta_0\as\xi)^k$.
These singularities must therefore be
resummed to all orders if one wishes the $x\to0$ limit to be
stable. 

\begin{figure}
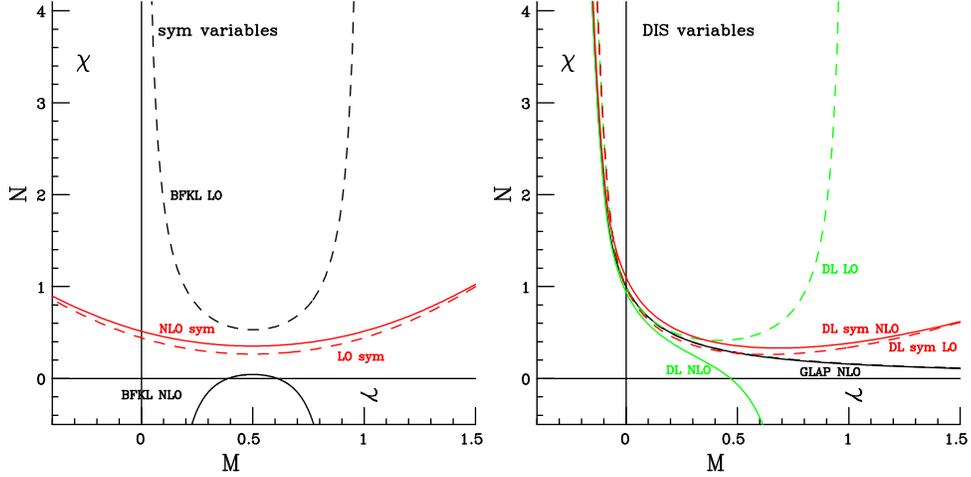

\centerline{\psfig{file=dualsyms.ps,width=0.5\linewidth}
\psfig{file=dualsym.ps,width=0.5\linewidth}}
%\vspace*{8pt}
\caption{Symmetrization of the $\chi$ kernel of Fig.~2.}
\vskip-0.5cm
\end{figure}
The resummation can be accomplished in an asymptotic
expansion,~\cite{Altarelli:2001ji,Altarelli:2003hk} 
it leads to the nonsingular result of Fig.~3, and it
has a further interesting consequence. Namely, if $\chi$ has a
minimum, duality implies that $\gamma$ has a cut: e.g. if $\chi$ is
quadratic, then $\gamma$ has a square-root cut. After running coupling
resummation, however, the cut in $\gamma$ is replaced by a simple pole
(Fig.~3, right). Interestingly, the pole is located on the real axis
to the left of the cut, as seen in Fig.~3 where a cut at $N\sim0.5$
leads to a pole at $N\sim 0.2$. Because the location $N_0$ of the rightmost
singularity in $\gamma$ implies a small $x$ behaviour of parton
distributions $\sim x^{-N_0}$ this means that running coupling
corrections considerably soften the resummed small $x$ behaviour, a
result first obtained in Ref.~\refcite{Ciafaloni:1999yw}.

\section{Gluon exchange symmetry}
The double-leading perturbative expansion of $\chi(M)$ is stable for
$M\lsim\frac{1}{2}$, but still unstable in the vicinity of $M=1$,
where it has alternating-sign poles. In particular, it will 
lack a minimum at even perturbative orders. This is problematic for
the running coupling resummation discussed in the previous section, 
which relies on the existence of a minimum.
The instability at $M=1$ can be understood~\cite{Fadin:1998py} and
cured~\cite{Salam:1998tj} on the basis of the observation that in
fact, due to the
symmetry
of the underlying Feynman diagrams upon interchange of incoming
and outgoing gluons, the BFKL kernel is symmetric about
$M=\frac{1}{2}$, i.e. $\chi(M)=\chi(1-M)$. This symmetry is broken by
the DIS choice of kinematical variables, which treats asymmetrically
the initial and final parton virtualities $\mu^2$ and $Q^2$, but it can be restored by
choosing e.g. $x=\frac{Q\mu}{s}$, where $s$ is the center-of-mass
energy of the partonic process. It is also broken by the asymmetric
choice of scale in 
running of the coupling $\as(Q^2)$. 

These symmetry breaking effects are 
computable and can be undone:~\cite{Altarelli:2004dq,Altarelli:2005xx} 
once the symmetry is restored, one can
symmetrize the DL expansion of $\chi$, thereby obtaining a kernel
which is perturbatively stable and free of poles at both $M=0$ and
$M=1$. In fact, the momentum conservation constraints implies that
symmetrized $\chi$ is an entire function of $M$, and has a minimum to
all orders. The symmetrized DL expansion of $\chi$ which ensues is
displayed in Fig.~4, where one sees that 
the powerful combination of duality and gluon symmetry
leads to a stable expansion: the LO and NLO
approximations are quite close. One can then revert to DIS variables
(Fig.~4, right),
thus obtaining the kernel which lead through duality and running
coupling resummation to an anomalous dimension and splitting function.
Interestingly,
symmetrization implies a further
softening of the kernel: the minimum of $\chi$, hence the branch cut of the
corresponding $\gamma$, is moved (for $\as=0.2$) from $N_0\sim0.5$ to
$N_0\sim0.3$. Note that the NLO resummed behaviour is now harder than
the LO one. 
\begin{figure}
\vskip-.6cm
\centerline{\psfig{file=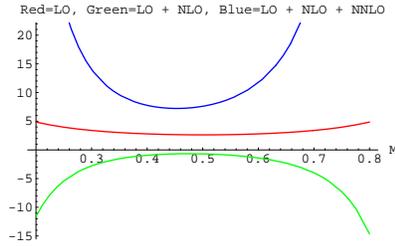,width=0.45\linewidth}
}
\vspace*{2pt}
\vskip-.4cm
\caption{Exact LO and NLO and approximate NNLO BFKL kernels with
  $\as=0.2$. Note the overall prefactor of $\as$ is omitted (unlike in
  previous Figs.~2--4).}
\vskip-.4cm
\end{figure}

The combination of symmetrization and running-coupling duality can be
exploited to obtain powerful analytic results. For instance, one can
use the knowledge of $\gamma$ up to
NNLO~\cite{Moch:2004pa,Vogt:2004mw} 
to determine~\cite{Ball:2005xx}  all singular
contributions to NNLO $\chi$, as shown in Fig.~6.
This requires a treatment of
running-coupling corrections up to NLO, and of various interference
terms, which is feasible if running coupling duality equations are
solved in an operator approach.~\cite{Ball:2005yy}

\section{Results}

\begin{figure}
\centerline{
\psfig{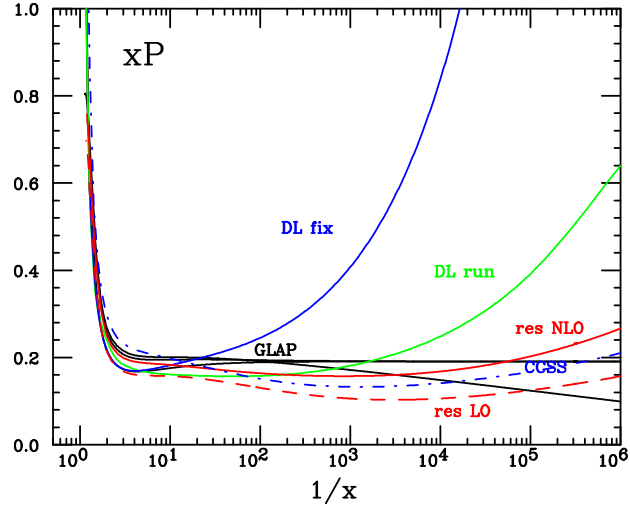}}
\vspace*{2pt}
\caption{The GLAP splitting function compared to various resummed approximations.}
\vskip-.3cm
\end{figure}
Various approximations to the  splitting function
are displayed in Fig.~6, determined with $n_f=0$ in order to avoid
problems related to the diagonalization of the splitting function
matrix.   The GLAP LO and NLO results are very close to each other and
coincide at small $x$ (their small difference is proportional to
$n_f$), but the NNLO GLAP result is seen to be unstable at small $x$
because of an unresummed $\frac{\as^3}{x}\ln\frac{1}{x}$  term (with
negative coefficient). The simple (LO) fixed-coupling double leading
result ('DL fix' in Fig.~6)
displays a dramatic rise $\sim x^{-0.5}$ which is certainly not seen in
the data. This is considerably softened by the running coupling
resummation (`DL run' in Fig.~6). A yet softer behaviour is obtained if the running
coupling resummation is applied to the symmetrized result (`res LO' and `res NLO' in Fig.~6), which also leads to a stable
perturbative expansion. 

For comparison the NLO curve (i.e. including NLO resummation of $\xi$
and $t$) obtained by
CCSS~\cite{Ciafaloni:2003ek,Ciafaloni:2005xx} 
is also shown.  The CCSS approach
starts from the BFKL equation, with the aim of determining the off-shell gluon
density, not just the anomalous dimensions of parton
distributions. This is then improved by enforcing symmetry, matching
to GLAP (analogous to duality), and running coupling. The main advantages of
this approach are its somewhat wider scope, and the fact that, being
based on the BFKL equation, it allows for an exact treatment of the
running coupling at small $x$ (unlike the ABF approach, where it is
determined through an asymptotic expansion). The main 
disadvantages are that, due to the latter feature, 
 anomalous
dimensions can only be obtained through a numerical deconvolution
procedure, and due to the former feature the perturbative expansion of the anomalous
dimension is not naturally organized as a leading log series (unlike
the DL approach), which implies for instance that the full NLO $n_f=4$
result in this approach is not yet available.

The NLO ABF and CCSS results turn out to be quite close: their
difference can be taken as an estimate of the intrinsic uncertainty in
the resummation procedure.
These results are now essentially ready for a phenomenological
implementation, though this will require the further development of
suitable tools, such as matched resummed coefficient functions (for
which all the required theoretical information is available), and
interpolation of the resummed results for their efficient numerical
implementation.

\section*{Acknowledgements}

I thank X.~Ji for inviting me to a stimulating meeting in a vibrant
city, and G.~Salam for many illuminating exchanges.


\begin{thebibliography}{0}

\bibitem{Lai:1999wy}
  H.~L.~Lai {\it et al.}  [CTEQ Collaboration],
  %``Global {QCD} analysis of parton structure of the nucleon: CTEQ5 parton
  %distributions,''
  Eur.\ Phys.\ J.\ C {\bf 12}, 375 (2000)
%  [arXiv:hep-ph/9903282].
  %%CITATION = HEP-PH 9903282;%%

\bibitem{Martin:2002aw}
  A.~D.~Martin, R.~G.~Roberts, W.~J.~Stirling and R.~S.~Thorne,
  %``Uncertainties of predictions from parton distributions. I: Experimental
  %errors. ((T)),''
  Eur.\ Phys.\ J.\ C {\bf 28}, 455 (2003)
 % [arXiv:hep-ph/0211080].
  %%CITATION = HEP-PH 0211080;%%

\bibitem{Ball:1995vc}
  R.~D.~Ball and S.~Forte,
  %``Summation of leading logarithms at small x,''
  Phys.\ Lett.\ B {\bf 351}, 313 (1995)
%  [arXiv:hep-ph/9501231].
  %%CITATION = HEP-PH 9501231;%%

\bibitem{Ball:1999sh}
  R.~D.~Ball and S.~Forte,
  %``The small x behaviour of Altarelli-Parisi splitting functions,''
  Phys.\ Lett.\ B {\bf 465}, 271 (1999)
%  [arXiv:hep-ph/9906222].
  %%CITATION = HEP-PH 9906222;%%


\bibitem{Altarelli:1999vw}
  G.~Altarelli, R.~D.~Ball and S.~Forte,
  %``Resummation of singlet parton evolution at small x,''
  Nucl.\ Phys.\ B {\bf 575}, 313 (2000)
  [arXiv:hep-ph/9911273].
  %%CITATION = HEP-PH 9911273;%%

\bibitem{Altarelli:2000mh}
  G.~Altarelli, R.~D.~Ball and S.~Forte,
  %``Small-x resummation and HERA structure function data,''
  Nucl.\ Phys.\ B {\bf 599}, 383 (2001)
 % [arXiv:hep-ph/0011270].
  %%CITATION = HEP-PH 0011270;%%

\bibitem{Altarelli:2001ji}
  G.~Altarelli, R.~D.~Ball and S.~Forte,
  %``Factorization and resummation of small x scaling violations with  running
  %coupling,''
  Nucl.\ Phys.\ B {\bf 621}, 359 (2002)
 % [arXiv:hep-ph/0109178].
  %%CITATION = HEP-PH 0109178;%%

\bibitem{Altarelli:2003hk}
  G.~Altarelli, R.~D.~Ball and S.~Forte,
  %``An anomalous dimension for small x evolution,''
  Nucl.\ Phys.\ B {\bf 674}, 459 (2003)
 % [arXiv:hep-ph/0306156].
  %%CITATION = HEP-PH 0306156;%%

\bibitem{Altarelli:2004dq}
  G.~Altarelli, R.~D.~Ball and S.~Forte,
  %``Progress in the small x resummation of the singlet anomalous dimension,''
  Nucl.\ Phys.\ Proc.\ Suppl.\  {\bf 135}, 163 (2004)
 % [arXiv:hep-ph/0407153].
  %%CITATION = HEP-PH 0407153;%%

\bibitem{Altarelli:2005xx}
  G.~Altarelli, R.~D.~Ball and S.~Forte, Preprint CERN-PH-TH/2005-174 

% No SPIRES record found for cite request Altarelli:2005xx

\bibitem{Salam:1998tj}
  G.~P.~Salam,
  %``A resummation of large sub-leading corrections at small x,''
  JHEP {\bf 9807}, 019 (1998)
 % [arXiv:hep-ph/9806482].
  %%CITATION = HEP-PH 9806482;%%

\bibitem{Ciafaloni:1999yw}
  M.~Ciafaloni, D.~Colferai and G.~P.~Salam,
  %``Renormalization group improved small-x equation,''
  Phys.\ Rev.\ D {\bf 60}, 114036 (1999)
 % [arXiv:hep-ph/9905566].
  %%CITATION = HEP-PH 9905566;%%

\bibitem{Ciafaloni:2003ek}
  M.~Ciafaloni, D.~Colferai, G.~P.~Salam and A.~M.~Stasto,
  %``Extending QCD perturbation theory to higher energies,''
  Phys.\ Lett.\ B {\bf 576}, 143 (2003)
 % [arXiv:hep-ph/0305254].
  %%CITATION = HEP-PH 0305254;%%

\bibitem{Ciafaloni:2003rd}
  M.~Ciafaloni, D.~Colferai, G.~P.~Salam and A.~M.~Stasto,
  %``Renormalisation group improved small-x Green's function,''
  Phys.\ Rev.\ D {\bf 68}, 114003 (2003)
 % [arXiv:hep-ph/0307188].
  %%CITATION = HEP-PH 0307188;%%

\bibitem{Thorne:2001nr}
  R.~S.~Thorne,
  %``The running coupling BFKL anomalous dimensions and splitting functions,''
  Phys.\ Rev.\ D {\bf 64}, 074005 (2001)
 % [arXiv:hep-ph/0103210].
  %%CITATION = HEP-PH 0103210;%%

\bibitem{Moch:2004pa}
  S.~Moch, J.~A.~M.~Vermaseren and A.~Vogt,
  %``The three-loop splitting functions in QCD: The non-singlet case,''
  Nucl.\ Phys.\ B {\bf 688}, 101 (2004)
 % [arXiv:hep-ph/0403192].
  %%CITATION = HEP-PH 0403192;%%

\bibitem{Vogt:2004mw}
  A.~Vogt, S.~Moch and J.~A.~M.~Vermaseren,
  %``The three-loop splitting functions in QCD: The singlet case,''
  Nucl.\ Phys.\ B {\bf 691}, 129 (2004)
 % [arXiv:hep-ph/0404111].
  %%CITATION = HEP-PH 0404111;%%

\bibitem{Moch:2005xu}
  S.~Moch, J.~A.~M.~Vermaseren and A.~Vogt,
  %``The longitudinal structure function at the third order,''
  Phys.\ Lett.\ B {\bf 606}, 123 (2005)
%  [arXiv:hep-ph/0411112].
  %%CITATION = HEP-PH 0411112;%%

\bibitem{Vermaseren:2005qc}
  J.~A.~M.~Vermaseren, A.~Vogt and S.~Moch,
  %``The third-order QCD corrections to deep-inelastic scattering by photon
  %exchange,''
  {\tt hep-ph/0504242.}
  %%CITATION = HEP-PH 0504242;%%

\bibitem{Martin:2004ir}
  A.~D.~Martin, R.~G.~Roberts, W.~J.~Stirling and R.~S.~Thorne,
  %``Physical gluons and high-E(T) jets,''
  Phys.\ Lett.\ B {\bf 604}, 61 (2004)
%  [arXiv:hep-ph/0410230].
  %%CITATION = HEP-PH 0410230;%%

\bibitem{Alekhin:2005gq}
  S.~Alekhin,
  %``Parton distribution functions from the precise NNLO QCD fit,''
  {\tt hep-ph/0508248.}
  %%CITATION = HEP-PH 0508248;%%

\bibitem{Thorne:2005da}
  R.~S.~Thorne, A.~D.~Martin, R.~G.~Roberts and W.~J.~Stirling,
  %``Recent progress in parton distributions and implications for LHC physics,''
  {\tt hep-ph/0507015.}
  %%CITATION = HEP-PH 0507015;%%
\bibitem{Blumlein:1997em}
  J.~Blumlein and A.~Vogt,
  %``The evolution of unpolarized singlet structure functions at small x,''
  Phys.\ Rev.\ D {\bf 58}, 014020 (1998)
%  [arXiv:hep-ph/9712546].
  %%CITATION = HEP-PH 9712546;%%

\bibitem{Ball:1997vf}
  R.~D.~Ball and S.~Forte,
  %``Asymptotically free partons at high energy,''
  Phys.\ Lett.\ B {\bf 405}, 317 (1997)
%  [arXiv:hep-ph/9703417].
  %%CITATION = HEP-PH 9703417;%%

\bibitem{Jaroszewicz:1982gr}
  T.~Jaroszewicz,
  %``Gluonic Regge Singularities And Anomalous Dimensions In QCD,''
  Phys.\ Lett.\ B {\bf 116}, 291 (1982).
  %%CITATION = PHLTA,B116,291;%%

\bibitem{Fadin:1998py}
  V.~S.~Fadin and L.~N.~Lipatov,
  %``BFKL pomeron in the next-to-leading approximation,''
  Phys.\ Lett.\ B {\bf 429}, 127 (1998)
%  [arXiv:hep-ph/9802290].
  %%CITATION = HEP-PH 9802290;%%
\bibitem{Ball:2005xx}
R.~D.~Ball, P.~Falgari, S.~Marzani and S.~Forte, {\it in preparation}

\bibitem{Ball:2005yy}
R.~D.~Ball and S.~Forte, {\it in preparation}

\bibitem{Ciafaloni:2005xx}
 M.~Ciafaloni, D.~Colferai, G.~P.~Salam and A.~M.~Stasto, 
in the Proceedings of the HERA-LHC workshop,
{\it in preparation}

% No SPIRES record found for cite request Ball:2005xx

% No SPIRES record found for cite request Ball:2005yy

% No SPIRES record found for cite request Ciafaloni:2005xx

\end{thebibliography}
\end{document}